\documentclass[12pt]{article}

\usepackage{scicite}

\usepackage{times}

\usepackage{graphicx}

\catcode`\@=11 
\def\lsim{\mathrel{\mathpalette\@versim<}}
\def\gsim{\mathrel{\mathpalette\@versim>}}
\def\@versim#1#2{\vcenter{\offinterlineskip
        \ialign{$\m@th#1\hfil##\hfil$\crcr#2\crcr\sim\crcr } }}

\newcommand{\arcsec}{$^{\prime\prime}$}
\newcommand{\arcmin}{$^{\prime}$}
\newdimen\sa  \newdimen\sb
\def\parcs{\sa=.07em \sb=.03em
     \ifmmode $\rlap{.}$^{\scriptscriptstyle\prime\kern -\sb\prime}$\kern -\sa$
     \else \rlap{.}$^{\scriptscriptstyle\prime\kern -\sb\prime}$\kern -\sa\fi}
\def\parcm{\sa=.08em \sb=.03em
     \ifmmode $\rlap{.}\kern\sa$^{\scriptscriptstyle\prime}$\kern-\sb$
     \else \rlap{.}\kern\sa$^{\scriptscriptstyle\prime}$\kern-\sb\fi}

\newenvironment{sciabstract}{%
\begin{quote} \bf}
{\end{quote}}

\newcounter{lastnote}
\newenvironment{scilastnote}{%
\setcounter{lastnote}{\value{enumiv}}%
\addtocounter{lastnote}{+1}%
\begin{list}%
{\arabic{lastnote}.}
{\setlength{\leftmargin}{.22in}}
{\setlength{\labelsep}{.5em}}}
{\end{list}}

\title{\bf A Dearth of Dark Matter in \\ Ordinary Elliptical Galaxies}

\author
{Aaron J. Romanowsky,$^{1,2\ast}$ Nigel G. Douglas,$^{2}$ Magda Arnaboldi,$^{3,4}$\\
 Konrad Kuijken,$^{5,2}$ Michael R. Merrifield,$^1$ Nicola R. Napolitano,$^2$\\
 Massimo Capaccioli,$^{3,6}$ Kenneth C. Freeman$^7$\\
\\
\normalsize{$^{1}$School of Physics \& Astronomy, University of Nottingham,}\\
\normalsize{University Park, Nottingham, NG7 2RD, England}\\
\normalsize{$^{2}$Kapteyn Astronomical Institute, Postbus 800, 9700 AV Groningen, The Netherlands}\\
\normalsize{$^3$INAF-Astronomical Observatory of Capodimonte,}\\ 
\normalsize{via Moiariello 16, I-80131 Naples, Italy}\\
\normalsize{$^4$INAF-Astronomical Observatory of Pino Torinese,}\\ 
\normalsize{via Osservatorio 20, I-10025 Pino Torinese, Italy}\\
\normalsize{$^5$Leiden Observatory, Postbus 9513, 2300 RA Leiden, The Netherlands}\\
\normalsize{$^6$Department of Physical Sciences, University ``Federico II'', Naples, Italy}\\
\normalsize{$^7$Research School of Astronomy \& Astrophysics, Mt. Stromlo Observatory,}\\
\normalsize{Cotter Road, Weston Creek, ACT 2611, Australia}\\
\\
\normalsize{$^\ast$To whom correspondence should be addressed;}\\ 
\normalsize{E-mail:  aaron.romanowsky@nottingham.ac.uk.}
}

\date{}

\begin{document} 




\maketitle


\begin{sciabstract}
The kinematics of the outer parts of three intermediate-luminosity
elliptical galaxies have been studied using the Planetary Nebula
Spectrograph.  The galaxies' velocity dispersion profiles are found to
decline with radius; dynamical modeling of the data indicates
the presence of little if any dark matter in these galaxies' halos.
This surprising result conflicts with findings in other galaxy types,
and poses a challenge to current galaxy formation theories.
\end{sciabstract}


Over the past twenty-five years, astronomers have gone from being
surprised by the existence of dark matter to the understanding that in
fact most of the Universe is dominated by exotic non-luminous
material.  In the prevailing paradigm, the gravitational influence of
``cold dark matter'' (CDM) is crucial to the formation of structure,
seeding the collapse and aggregation of today's luminous
systems.  An inherent consequence of this picture is that galaxies
have massive, extended CDM halos.  Indeed, such halos are evident around spiral galaxies,
where the rotational speeds in their extended cold gas disks do not
decrease outside the visible stars---a
gravitational signature of dark matter\cite{persic96}.

The evidence for dark matter in elliptical galaxies is still circumstantial.
Assessments of the total masses of individual
elliptical systems have generally been confined to the very brightest
ones, where the gravitational potential may be measured using
x-ray emission\cite{loewenstein99} or strong gravitational
lensing\cite{keeton01}, and to nearby dwarfs, where the
kinematics of individual stars offer a probe of the mass
distribution\cite{kleyna02}.  More ``ordinary'' elliptical galaxies,
with luminosities close to the characteristic $L^*$ (= 2.2$\times
10^{10} L_{B,\odot}$ in $B$-band solar units for a Hubble constant
$H_0=70$~km~s$^{-1}$~Mpc$^{-1}$)\cite{madgwick02}, are more
difficult to study because in general they lack a
simple kinematical probe at the larger radii where dark matter is
expected to dominate.  The velocity distribution of the diffuse stellar
light is the natural candidate\cite{kronawitter00}, but studies have
been limited by the faintness of galaxies' outer parts to radii
$\lsim$~2$R_{\rm eff}$ (where $R_{\rm eff}$ is the galaxy's
``effective radius'', enclosing half its projected light).

A powerful alternative is offered by planetary nebulae (PNe), which
are detectable even in distant galaxies through their characteristic
strong emission lines; once found, their line-of-sight velocities can
then be readily determined by the Doppler shift in these lines.  These
objects have been used in the past as tracers of the stellar
kinematics of galaxies\cite{hui95}, but the procedure of locating them
using narrow-band imaging surveys and then blindly obtaining spectra
at the identified positions has proved difficult to implement
efficiently on a large scale.  We have therefore developed a
specialized instrument, the Planetary Nebula Spectrograph (PN.S),
specifically to study the kinematics of PNe in elliptical
galaxies\cite{douglas02}.  The PN.S uses counter-dispersed imaging (a
type of slitless spectroscopy) over a wide field to simultaneously
detect and measure velocities for PNe using their [O III] emission at
500.7~nm.  Its optimization for this purpose means that the PN.S is far
more efficient for extragalactic PN studies than any other existing instrumentation.

Observations with the PN.S on the 4.2-m William Herschel Telescope
have allowed us to extend stellar kinematic studies to the outer parts
of three intermediate-luminosity elliptical galaxies, NGC 821, NGC
3379, and NGC 4494 (Table 1).  In each of these systems, we have
measured $\sim$~100 PN velocities with uncertainties of
20~km~s$^{-1}$ out to radii of 4--6$R_{\rm eff}$ (Figs.~1--3).
The line-of-sight velocities in the outer parts of all these galaxies 
show a clear decline in dispersion with radius (Fig.~4).
A decrease in random velocities with radius has been
indicated by small samples of PNe around NGC~3379\cite{ciardullo93},
but the more extensive data set presented
here provides a more definitive measurement of this decline, and
reveals that it also occurs in other similar galaxies.  The new data
are inconsistent with simple dark halo models (Fig.~4) and
thus different from kinematical results for
brighter ellipticals [e.g., \cite{kronawitter00,magorrian01}].

More surprisingly, the velocity dispersion data follow
simple models containing no dark matter (Fig.~4), showing
the nearly Keplerian decline with radius outside 2~$R_{\rm eff}$
that such models predict, and
suggesting that these systems are not embedded in massive
dark halos.  However, a declining profile can also be the signature of
a distribution of orbits that is primarily radial, because the observable
line-of-sight component of velocity in such a system will decline with
radius even if the magnitude of the velocity remains constant due to a
surrounding dark halo.  To assess the impact of this 
well-known degeneracy between mass and orbital anisotropy, we have
constructed spherical Jeans models (see Appendix) with a fixed anisotropy, spanning a range
from radially- to tangentially-biased extremes.  We have fitted these models
to the observed velocity dispersion profiles for the
stars\cite{jedrzejewski89,statler99,bender94} and the PNe, inferring a benchmark quantity 
$\Upsilon_{B\rm 5}$, the $B$-band mass-to-light ratio in solar units
at 5~$R_{\rm eff}$ (Table 1).  Even with this extra degree of freedom
in the modeling, the values of 5--17 are low compared with the typical
$\Upsilon_{B\rm 5}$ of $\sim$~20--40 found for bright elliptical
galaxies\cite{loewenstein99,bahcall95}.

These models are not completely definitive, as there
is no physical reason why the orbital anisotropy should not vary with
radius.  But we also have not used all the information contained
in the kinematical data, fitting only binned root-mean-square values of
the stellar and PN velocities, rather than the full two-dimensional distribution
of line-of-sight velocities as a function of projected radius.
To overcome these remaining limitations, we have used
an orbit library method (see App.) to model the dynamics of NGC~3379.
This procedure takes a broad suite of parameterized mass models,
allows for orbital anisotropies that vary arbitrarily with radius,
and finds the best fit to
the PN velocity data as well as to the stellar brightness distribution\cite{capaccioli90} 
and kinematics\cite{statler99},
including the higher-order velocity moments that quantify the shape 
of the velocity distribution.

This procedure's only underlying assumptions are that Newtonian gravity applies,
that the galaxy is in equilibrium, and that it is spherically symmetric.
With respect to the latter assumption, 
the observed ellipticity of NGC~3379 varies from 0.14 to 0.30,
so that its gravitational potential can be approximated as spherical to better than 10\% accuracy.
More serious is the possibility that this system contains a face-on disk component
similar to an S0 galaxy, as has been suggested for faint ellipticals in general\cite{rix99}
and for NGC~3379 in particular\cite{capaccioli90};
such a cold component could be responsible for the apparent low velocities.
However, kinematical modeling of the inner regions\cite{statler01}
and the weak rotation of the outer parts indicate that this is unlikely;
furthermore, the galaxy's globular cluster system also shows a steeply declining
dispersion profile\cite{bridges03}, and this population is unlikely to contain
a disk component.  The model also relies on the PNe following the same spatial
and velocity distribution as the main stellar component;
this is not strictly true because oxygen abundance radial gradients
make the PN detectability vary with radius,
but we calculate that this will cause a mass underestimate of only $\sim$~5\%.
Another potential complication is that background emission-line galaxies can
masquerade as PNe. We have discarded several of these from the data set
on the basis of their extended appearance, and one that is a $>$3-$\sigma$ outlier;
any remaining contaminants would cause an apparent increase in the
velocity dispersion at large radii, which is not seen in the data
so the sample does not appear to be significantly contaminated.
Within these assumptions, the modeling process 
should uncover the range of possible combinations of mass and orbit distribution that
could reproduce the galaxy's observed properties. 

\newpage
Using this method to analyze the data for NGC~3379 (Fig.~A1) reveals that
$\Upsilon_{B5}$ is tightly constrained to be only $6.4\pm0.6$.  Because
population synthesis models indicate that the stars of NGC~3379 have
$\Upsilon_{B\star}=$~4--9\cite{gerhard01}, 
and dynamical models of the inner regions where little dark matter is expected
indicate $\Upsilon_{B\star}=$6--7\cite{gebhardt00},
this result suggests a ``naked'' galaxy with no dark matter cloaking its visible body.
However, the mass distribution inferred by this modeling process does
not quite match the light distribution (Fig.~A2),
so there appears to be a small amount of dark matter.
A dark matter-less galaxy is still possible if the
baryonic $\Upsilon_B$ varies somewhat with radius, due to a gradient
in the properties of the stellar population, or due to the presence of
some undetectable baryonic component such as warm hydrogen gas.
Such models may be plausible but are also somewhat {\it ad hoc}.  In any case, 
the ratio of dark matter to total mass within 5~$R_{\rm eff}$ (9 kpc) is
$\leq$~0.32, so the galaxy within
this region is not dark matter dominated.  Extrapolating
the models to the virial radius (120~kpc), we find
$\Upsilon_B=32\pm13$, which is consistent with the value of
$\Upsilon_B=27\pm5$ found by an independent analysis of
the kinematics of a cold gas ring at 110~kpc\cite{schneider89}.

Out of the five intermediate-luminosity
ellipticals studied to date with extended kinematics [the three from
this study as well as NGC 2434\cite{rix97} and NGC 4697\cite{mendez01}],
four have declining velocity dispersions.  The fourth, NGC~4697, has
$\Upsilon_{B5}$~$\sim$~11 from Jeans modeling, so all four galaxies
are consistent with $\Upsilon_{B5}$~$\lsim$~13.  
The apparent low dark matter content in these systems could be explained by
pathological orbit structures (such as rapid strong anisotropy variations with radius),
but the orbit library analysis rules out this possibility for NGC~3379,
and in general such behavior has not been found in similar detailed modeling of elliptical galaxies.
Latent disks could be to blame,
but as with NGC~3379, the observed rotational speeds are low, and hiding them adequately
would require coincidentally unpropitious viewing angles.
Thus it seems most plausible that the apparent Keplerian decline in the velocity
dispersions of all these systems is what it seems to be,
and many ordinary elliptical galaxies are 
highly deficient in dark matter relative to other galaxy types---a
possibility already hinted at
by previous dynamical studies\cite{bertin94,gerhard01,capaccioli03}.

This result clashes with conventional conceptions of galaxy formation: in
particular, if ellipticals are built up by mergers of smaller 
galaxies, it is puzzling for the resulting
systems to show little trace of their precursors' dark matter halos.
More detailed comparisons with the predictions of the standard CDM paradigm
are not yet possible, because
the baryonic processes during galaxy formation are complex, and
high-resolution {\it ab initio} simulations are so far unable to
reproduce an elliptical galaxy.
However, estimating the effects of baryonic collapse on a CDM
halo\cite{navarro97} using an adiabatic approximation yields
$\Upsilon_{B5}\sim$~20.  Low resolution simulations including baryonic
processes\cite{weinberg03}, extrapolated inward in radius, also
predict $\Upsilon_{B5}$~$\sim$~20.  
These values conflict with those derived for the observed galaxies above.
On the other hand, CDM predictions at the virial radius are generally supported by 
statistical studies of $L^*$ ellipticals using weak gravitational 
lensing\cite{wilson01}.  This apparent disparity could be resolved if
the dark matter content of ellipticals varies, or
if these galaxies have large amounts of dark matter spread 
out to still larger radii than the PNe can probe.
The former scenario begs a suitable mechanism for dark matter depletion,
and the latter violates CDM predictions for high central concentrations of
dark halos---a problem that is also increasingly evident in other stellar systems\cite{ostriker03}.

It is apparent that some important physics is still missing from the recipes
for galaxy formation.  One obvious candidate is that
substantial portions of these galaxies' dark matter halos have been
shed through interactions with other galaxies.  Such stripping has been
inferred for ellipticals near the centers of dense galaxy
clusters\cite{natarajan02}, but the galaxies we studied here are in much
sparser environments, where substantial stripping 
is not expected to have been an important process.
Crucial to understanding the incidence and origin of this low dark matter phenomenon
will be results for a large sample of ellipticals with a broad range of properties, 
including differing environmental densities, which could be a key factor
in determining halo outcomes; the continuing PN.S observing program will provide this sample.


\bibliography{scibib}

\begin{thebibliography}{}

\bibitem {persic96} M. Persic, P. Salucci, F. Stel, {\it Mon. Not. R. Astron. Soc.} {\bf 281}, 27 (1996).

\bibitem {loewenstein99} M. Loewenstein, R. E. White III.,
  {\it Astrophys. J.} {\bf 518}, 50 (1999).

\bibitem {keeton01} C. R. Keeton, {\it Astrophys. J.} {\bf 561}, 46 (2001).

\bibitem {kleyna02} J. Kleyna, M. I. Wilkinson, N. W. Evans, G. Gilmore, C. Frayn,
  {\it Mon. Not. R. Astron. Soc.} {\bf 330}, 792 (2002).

\bibitem {madgwick02} D. S. Madgwick {\it et al.}, {\it Mon. Not. R. Astron. Soc.} {\bf 333}, 133 (2002).

\bibitem {kronawitter00} A. Kronawitter, R. P. Saglia, O. Gerhard, R. Bender, {\it Astron. Astrophys.  Suppl. Ser.} {\bf 144}, 53 (2000).

\bibitem {hui95} X. Hui, H. C. Ford, K. C. Freeman, M. A. Dopita, {\it Astrophys. J.} {\bf 449}, 592 (1995).

\bibitem {douglas02} N. G. Douglas {\it et al.}, {\it Publ. Astron. Soc. Pac.} {\bf 114}, 1234 (2002).

\bibitem {ciardullo93} R. Ciardullo, G. H. Jacoby, H. B. Dejonghe, {\it Astrophys. J.} {\bf 414}, 454 (1993).

\bibitem {magorrian01} J. Magorrian, D. Ballantyne, {\it Mon. Not. R. Astron. Soc.} {\bf 322}, 702 (2001).

\bibitem {jedrzejewski89} R. Jedrzejewski, P. L. Schechter, {\it Astron. J.} {\bf 98}, 147 (1989).

\bibitem {statler99} T. S. Statler, T. Smecker-Hane, {\it Astrophys. J.} {\bf 117}, 839 (1999).

\bibitem {bender94} R. Bender, R. P. Saglia, O. E. Gerhard, {\it Mon. Not. R. Astron. Soc.} {\bf 269}, 785 (1994).

\bibitem {bahcall95} N. A. Bahcall, L. M. Lubin, {\it Astrophys. J.} {\bf 447}, L81 (1995).

\bibitem {capaccioli90} M. Capaccioli, E. V. Held, H. Lorenz, M. Vietri, {\it Astron. J.} {\bf 99}, 1813 (1990).

\bibitem {rix99} H.-W. Rix, C. M. Carollo, K. Freeman, {\it Astrophys. J.} {\bf 513}, L25 (1999).

\bibitem {statler01} T. S. Statler, {\it Astron. J.} {\bf 121}, 244 (2001).

\bibitem {bridges03} M. A. Beasley, T. J. Bridges, D. A. Forbes, in preparation (2003).

\bibitem {gerhard01} O. Gerhard, A. Kronawitter, R. P. Saglia, R. Bender, {\it Astronom. J.} {\bf 121}, 1936 (2001).

\bibitem {gebhardt00} K. Gebhardt {\it et al.}, {\it Astronom. J.} {\bf 119}, 1157 (2000).

\bibitem {schneider89} S. E. Schneider, {\it et al.}, {\it Astronom. J.} {\bf 97}, 666 (1989).

\bibitem {rix97} H.-W. Rix, P. T. de Zeeuw, N. Cretton, R. P. van der Marel, C. M. Carollo, {\it Astrophys. J.} {\bf 488}, 702 (1997).

\bibitem {mendez01} R. H. M\'{e}ndez {\it et al.}, {\it Astrophys. J.} {\bf 563}, 135 (2001).

\bibitem {bertin94} G. Bertin {\it et al.}, {\it Astron. Astrophys.} {\bf 292}, 381 (1994).

\bibitem {capaccioli03} M. Capaccioli, N. R. Napolitano, M. Arnaboldi, in {\it Proc. Sakharov Conf. of Physics}, in press (available at http://xxx.lanl.gov/abs/astro-ph/0211323).

\bibitem {navarro97} J. F. Navarro, C. S. Frenk, S. D. M. White, {\it Astrophys. J.} {\bf 490}, 493 (1997).

\bibitem {weinberg03} D. H. Weinberg, R. Dav\'{e}, N. Katz, L. Hernquist, {\it Astrophys. J.}, submitted (available at http://xxx.lanl.gov/abs/astro-ph/0212356).

\bibitem {wilson01} G. Wilson, N. Kaiser, G. A. Luppino, L. L. Cowie, {\it Astrophys. J.} {\bf 555}, 572 (2001).

\bibitem {ostriker03} J. P. Ostriker, P. Steinhardt, {\it Science} {\bf 300}, 1909 (2003).

\bibitem {natarajan02} P. Natarajan, J.-P. Kneib, I. Smail, {\it Astrophys. J.} {\bf 580}, L11 (2002).

\bibitem {tonry01} J. L. Tonry {\it et al.}, {\it Astrophys. J.} {\bf 546}, 681 (2001).

\bibitem {RC3} G. de Vaucouleurs, {\it et al.},
{\it Third Reference Catalogue of Bright Galaxies} (Springer-Verlag, New York, 1991).

\bibitem {faber89} S. M. Faber, {\it et al.}, {\it Astrophys. J. Suppl. Ser.} {\bf 69}, 763 (1989).

\bibitem {jacoby96} G. H. Jacoby, R. Ciardullo, W. E. Harris, {\it Astrophys. J.} {\bf 462}, 1 (1996).

\bibitem {dss} The compressed files of the ``Palomar Observatory - Space Telescope
Science Institute Digital Sky Survey'' of the northern sky, based on
scans of the Second Palomar Sky Survey are copyright (c) 1993-2000 by the California Institute of Technology.
Produced under Contract No. NAS5-2555 with the National Aeronautics and Space Administration.

\bibitem {binney87} J. Binney, S. Tremaine, {\it Galactic Dynamics} (Princeton Univ. Press, Princeton, PA, 1987), sec. 4.2.

\bibitem {hernquist90} L. Hernquist, {\it Astrophys. J.} {\bf 356}, 359 (1990).

\bibitem {romanowsky01} A. J. Romanowsky, C. S. Kochanek, {\it Astrophys. J.} {\bf 553}, 722 (2001).



\end{thebibliography}

\bibliographystyle{Science}


\begin{scilastnote}
\item Based on observations made with the William Herschel Telescope operated
    on the island of La Palma by the Isaac Newton Group in the Spanish
    Observatorio del Roque de los Muchachos of the Instituto de
    Astrofisica de Canarias.
Use was made of the HyperLeda galaxy database: 
http://www-obs.univ-lyon1.fr/hypercat/ .
    We thank Ortwin Gerhard and the anonymous referee for their
    helpful comments for improving the paper.
NRN is supported by the European Commision FP5 Marie Curie Fellowships Programme.

\end{scilastnote}

\vskip 0.5cm


\begin{table}
\begin{tabular}{lccccccc}
{ Galaxy} & { Type} & { $D$ (Mpc)} & $R_{\rm eff}$ & $M_B$ & { Observations} & {  $N$(PNe)} & $\Upsilon_{B5}$ ($\Upsilon_{B,\odot}$)\\
\cline{1-8} \\
\multicolumn{1}{l}{NGC 821} & E2 & 24 & 50\arcsec & -20.5 & Sep 01, 11h  & 104 & 13--17 \\
\multicolumn{1}{l}{NGC 3379} & E1 & 11 & 35\arcsec & -20.0 & Mar 02, 3h & 109 & 5--8 \\
\multicolumn{1}{l}{NGC 4494} & E0 & 17 & 49\arcsec & -20.6 & Mar 02, 4h & 73 & 5--7
\end{tabular}
\caption{
  Properties of three galaxies observed with the Planetary Nebula Spectrograph.
  Columns show galaxy name, type, distance\cite{tonry01}, effective radius\cite{RC3},
  absolute magnitude, date and effective
  integration time of observations (normalized to 1\arcsec{} seeing),
  number of PNe detected,
  and the range of {\it B}-band mass-to-light ratio at 5 effective radii (in solar units) 
  determined from anisotropic Jeans models.
  The mass-to-light ratios are systematically sensitive to the assumed distance---e.g.,
  if $D=$~33~Mpc for NGC~821\cite{faber89},
  $\Upsilon_{B5}$ is decreased by 30\%;
  and if $D =$~11--13~Mpc for NGC~4494\cite{faber89,jacoby96},
  $\Upsilon_{B5}$ is increased by 30--60\%.
  The characteristic magnitude is $M_B^*$~=~-20.4\cite{madgwick02}.
  }
\end{table}

\clearpage

\includegraphics[width=\textwidth]{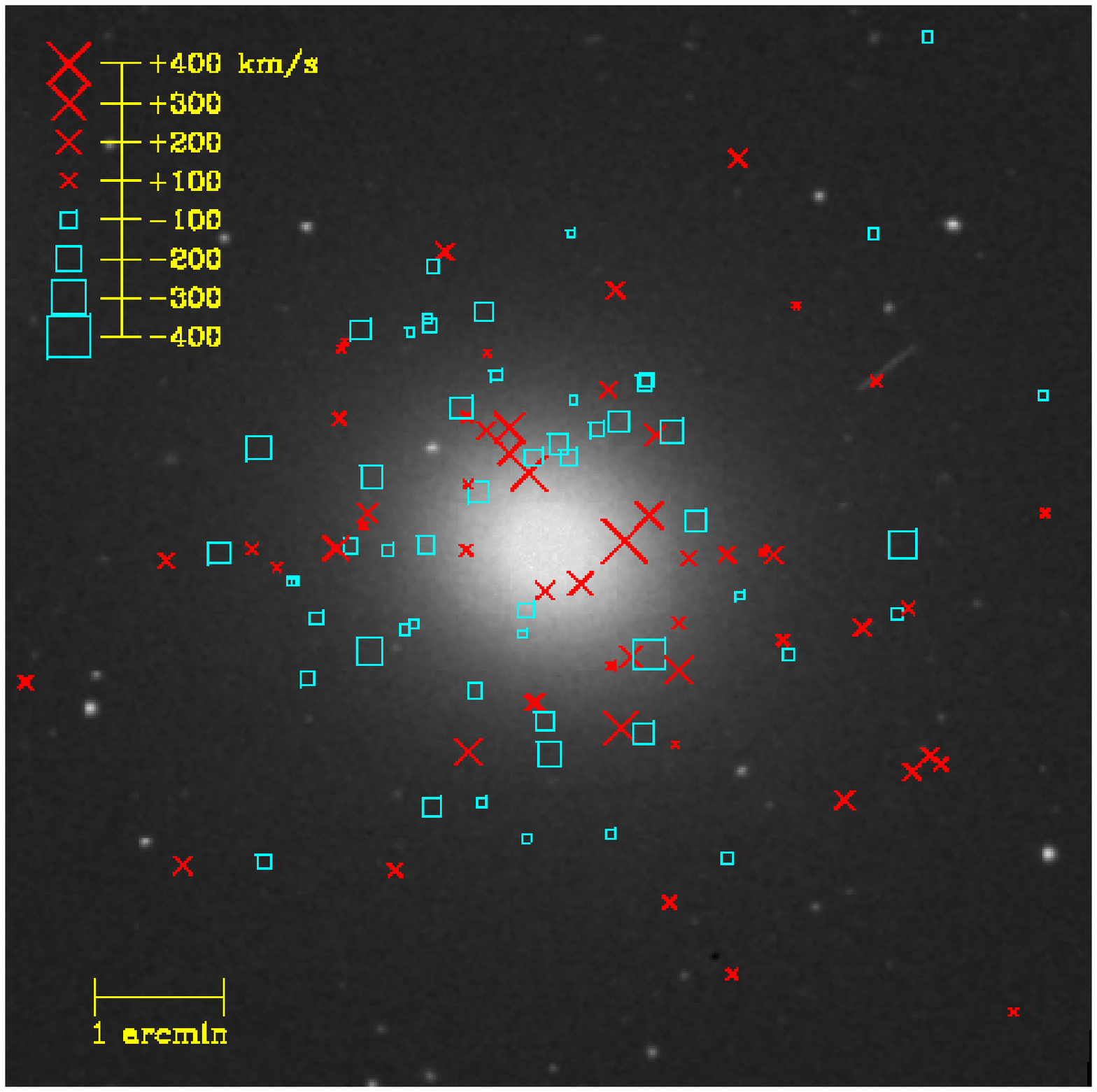}

\vskip 1.5cm
\noindent {\bf Fig. 1.}
NGC 3379 with
109 planetary nebula line-of-sight velocities
relative to the systemic velocity,
as measured with the William Herschel 4.2-m telescope
and the PN.S instrument.
Red crosses represent receding velocities,
and blue boxes are approaching,
where the symbol sizes are proportional to the velocity magnitudes.
The background image is from the Digitized Sky Survey\cite{dss}, and
the field of view is 8\parcm4$\times$8\parcm4 = 26$\times$26 kpc = 14$\times$14~$R_{\rm eff}$.

\newpage
\includegraphics[width=\textwidth]{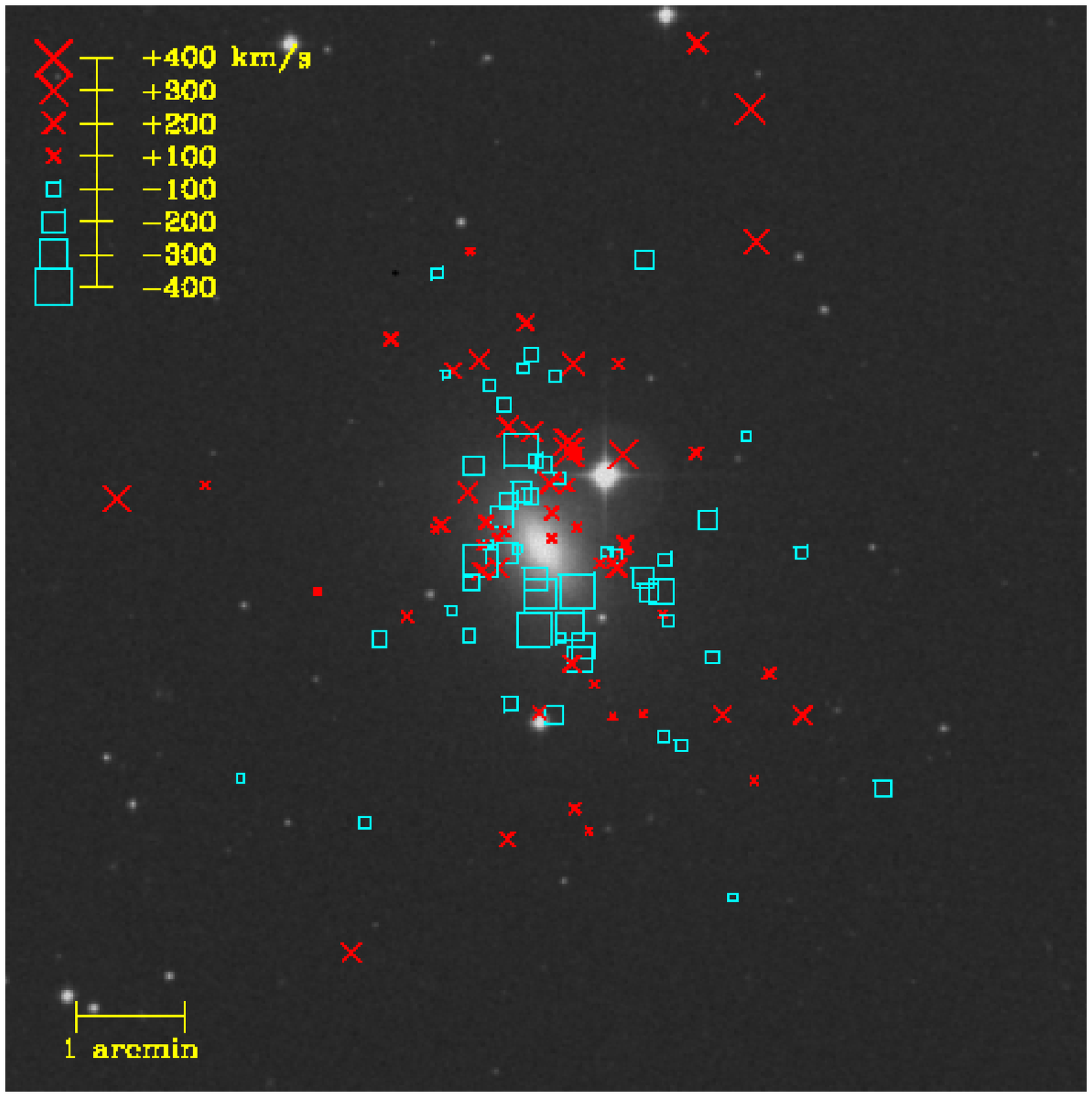}

\vskip 1.5cm
\noindent {\bf Fig. 2.}
NGC~821 with 104 planetary nebula velocities.
See Fig. 1 for further description.
The field of view is 10\arcmin$\times$10\arcmin = 70$\times$70 kpc = 12$\times$12~$R_{\rm eff}$.

\newpage
\includegraphics[width=\textwidth]{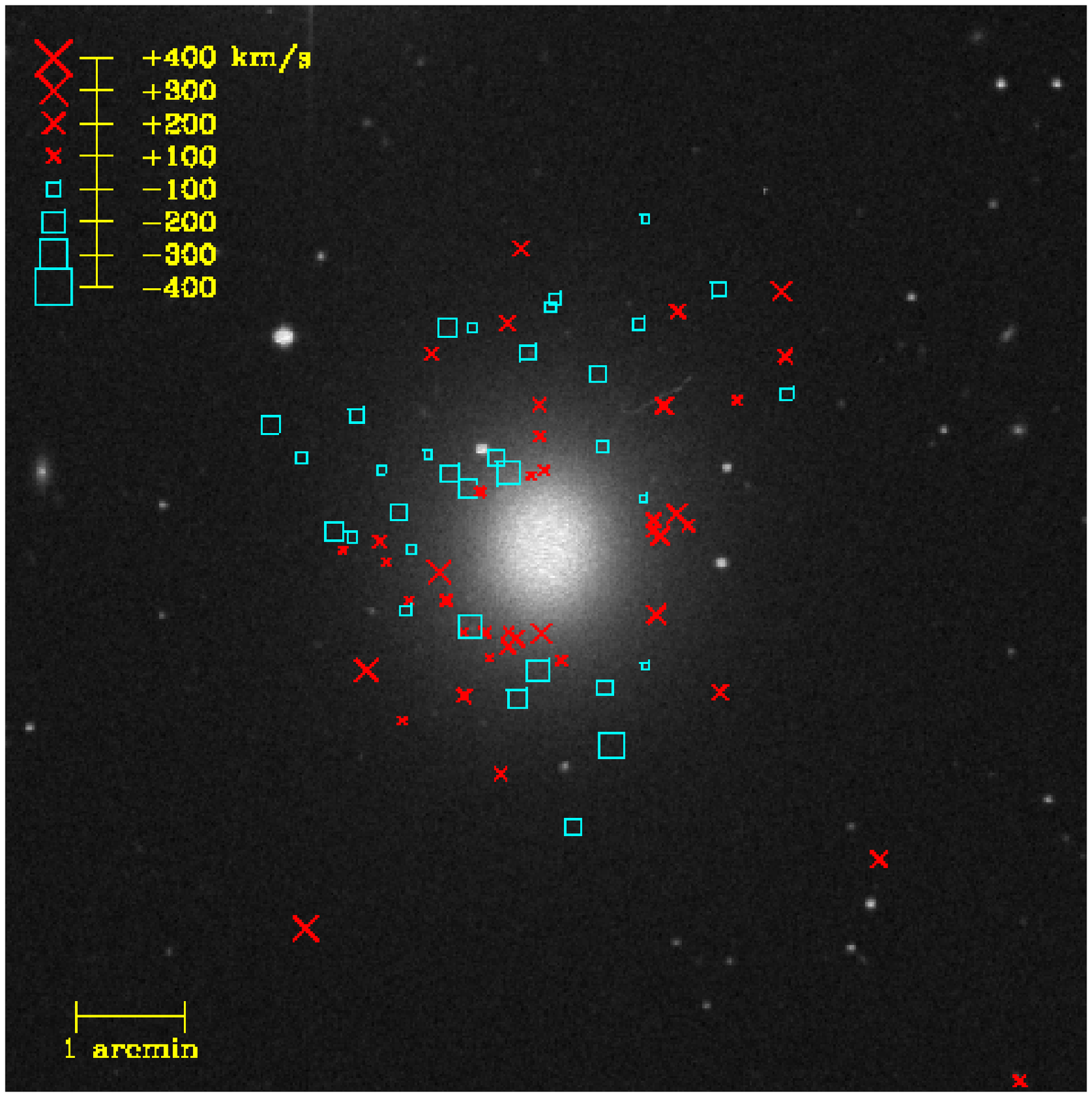}

\vskip 1.5cm
\noindent {\bf Fig. 3.}
NGC~4494 with 73 planetary nebula velocities.
See Fig. 1 for further description.
The field of view is 10\arcmin$\times$10\arcmin = 50$\times$50 kpc = 12$\times$12~$R_{\rm eff}$.

 \includegraphics[width=\textwidth]{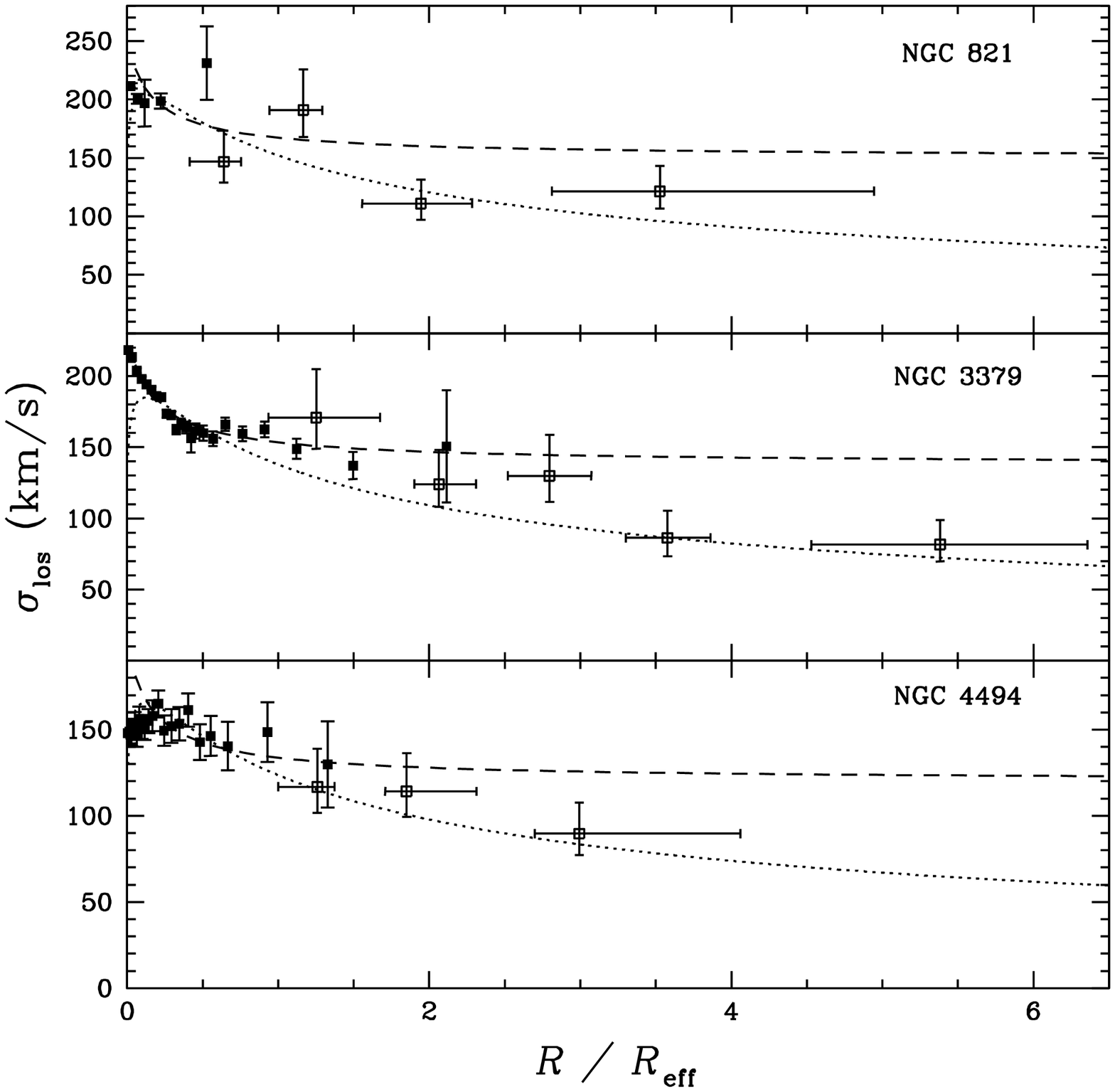}
\noindent {\bf Fig. 4.}
  Line-of-sight velocity dispersion profiles for three elliptical galaxies,
  as a function of projected radius in units of the effective radius.
  Open points show planetary nebula data (from PN.S);
  solid points show diffuse stellar data\cite{jedrzejewski89,statler99,bender94}.
  The vertical error bars show 1~$\sigma$ uncertainties in the dispersion,
  and the horizontal error bars show the radial range covered by 68\% of the points in each bin.
  Predictions of simple isotropic models are also shown for comparison:  
  a singular isothermal halo (dashed lines) and a constant mass-to-light ratio galaxy (dotted lines).

\newpage

\section*{Appendix: Methods}


\subsection*{Jeans models}

The non-rotating spherical Jeans equations link the second moments $\sigma_i(r)$ of a system's velocity
distribution to its mass $M(r)$ and luminosity $\nu(r)$ distributions\cite{binney87}.
We use the formulation
\begin{equation}
M(r) = - \frac{r \sigma_r^2}{G}  \left(\frac{d\ln \nu}{d\ln r} + \frac{d\ln \sigma^2_r}{d\ln r} + 2\beta\right) ,
\end{equation}
where $\sigma_r^2$ and $\sigma_{\theta}^2$ are the radial and tangential velocity dispersions,
and $\beta \equiv 1 - \sigma_{\theta}^2/\sigma_r^2$ quantifies the
(unknown) velocity anisotropy,
We take the luminosity distribution to be the Hernquist\cite{hernquist90} profile 
$\nu(r)=L (2\pi r)^{-1} (r+a)^{-3}$ corresponding to the known effective radius
of the galaxy $(R_{\rm eff} = 1.8153 a)$,
and parametrize the dispersion profile by the form
\begin{equation}
\sigma_r^2 = v_0^2 \left[1 + \left(\frac{r}{r_0}\right)^{\alpha \delta}\right]^{-1/\delta} 
\end{equation}
(thus allowing for a dark halo with a range of radial mass distributions),
which is projected and fitted to the dispersion data
while optimizing for the parameters \{$v_0,r_0,\alpha,\delta$\}.
The anisotropy is set to be constant and is varied between the
reasonable extremes of $\beta=-0.5$ (tangential) to $\beta=0.5$ (radial).
We then solve the Jeans equation for the mass profile $M(r)$.


These models are suited to exploring the range of outer galaxy mass profiles,
and not for reproducing the fine structure of the dispersion data inside 1~$R_{\rm eff}$.
They are thus illustrative of the range of $\Upsilon_{B5}$
associated with the anisotropy uncertainty,
and these are the $\Upsilon_{B5}$ values reported in Table 1;
the statistical uncertainty is of the same order of magnitude as
this systematic uncertainty.
As discussed for NGC~3379, the assumption of spherical symmetry
should be accurate enough for mass estimates at the 10\% level.

\subsection*{Orbit models}

Our orbit library method is documented in \cite{romanowsky01}.
In summary, it begins with an assumed form for the mass distribution,
and in its gravitational potential calculates a library of orbits,
of all types from radial to circular.
These individual orbits are then combined non-parametrically to find the best fit to
the observational constraints.
Regularization is used to stabilize the solution, but at such a low level as to avoid biasing them.
The fit to the data is quantified by a likelihood function ${\cal L}$, and 68\% confidence limits in
model parameters are given by $\Delta \ln {\cal L}=1/2$.
Because the degrees of freedom are ill-defined with this method,
the absolute value of $\ln {\cal L}$ is not very informative, but
the diffuse stellar component of the fits (photometric and kinematic) has a typical 
$\chi^2=8$--$9$ for 129 data points,
and the models qualitatively reproduce the data properties.


We use a grid of models consisting of a constant mass-to-light ratio 
Hernquist model galaxy\cite{hernquist90} plus a CDM simulated dark halo\cite{navarro97},
with the mass density distribution:
\begin{equation}
\rho(r) = \frac{v_*^2 a^2}{2 \pi G r (r+a)^{3}} +
\frac{v_s^2 r_s}{4 \pi G r(r_s+r)^{2}} .
\end{equation}
The mass model parameters are thus the halo core radius $r_s$, 
the mass of the stars $v_*^2$, and
the relative masses of the stars and halo $v_s^2/v_*^2$.
Our grid covers values of $r_s$ from 113\arcsec {} to 700\arcsec {} (6 to 36 kpc),
$v_*$ from 575 to 625 km~s$^{-1}$, and
$v_s^2/v_*^2$ from 0 to 1.1.


The best-fit solution for NGC~3379 is shown in Fig.~A1,
and the range of permitted solutions is illustrated by Fig.~A2.
We find $\Upsilon_{B5}$ (in solar units) to be
$6.4\pm0.6$ at 185\arcsec {} (9 kpc),
$7.9\pm1.1$ at 270\arcsec {} (14 kpc),
and $32\pm13$ at 2460\arcsec {} (120 kpc).
Allowable solutions within the 68\% confidence limits include
\{$r_s$, $v_*$, $v_s^2/v_*^2$\} = 
\{199\arcsec{}, 603, 0.15\},
\{280\arcsec{}, 610, 0.15\},
\{350\arcsec{}, 610, 0.12\},
\{350\arcsec{}, 603, 0.20\},
\{500\arcsec{}, 610, 0.12\},
\{500\arcsec{}, 603, 0.20\},
and
\{500\arcsec{}, 603, 0.30\},
where $v_*$ is in km~s$^{-1}$.

\newpage
 \includegraphics[width=\textwidth]{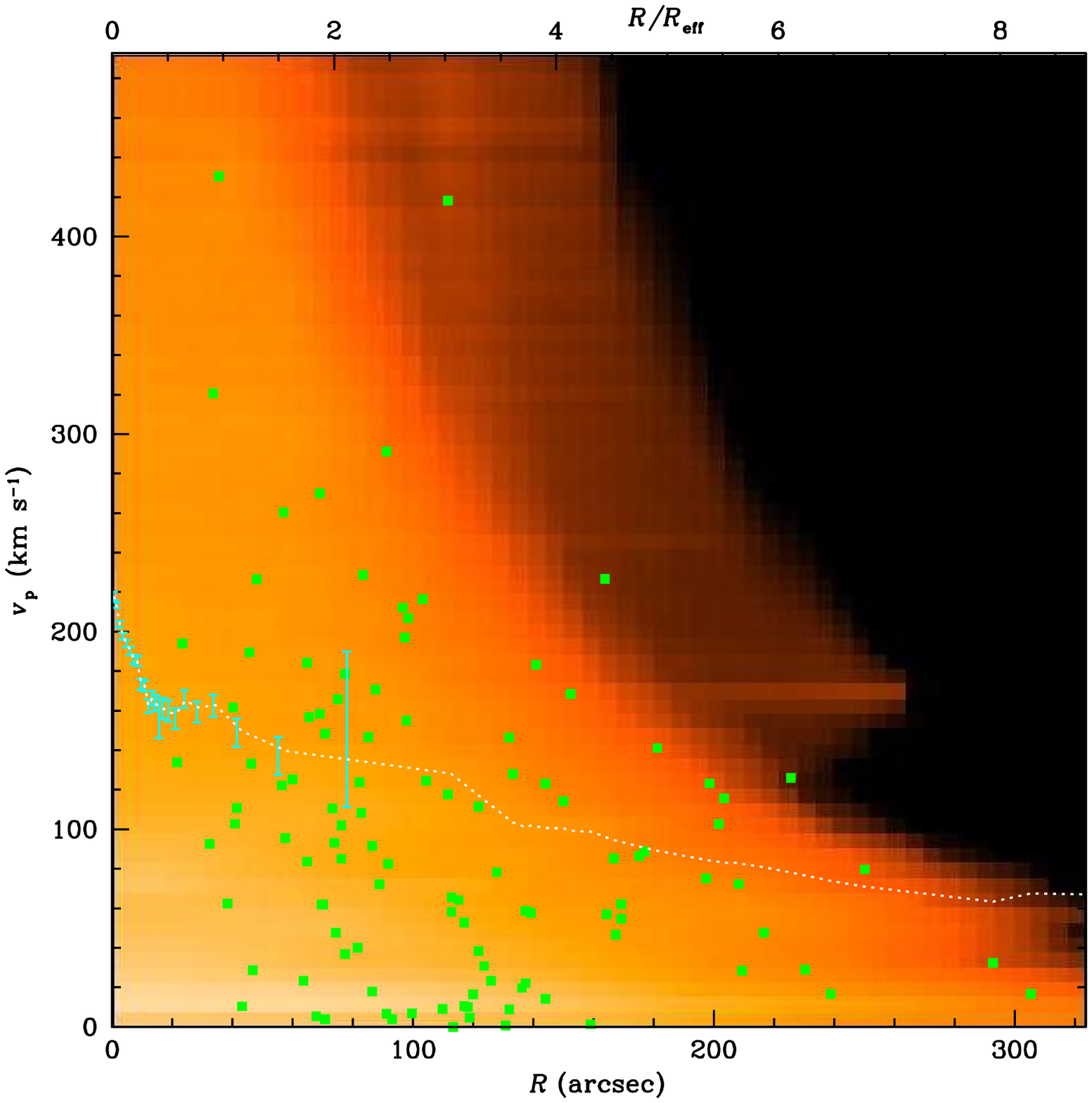}
\noindent {\bf Fig. A1.}
Line-of-sight velocity distribution with radius in NGC~3379.
The error bars show stellar dispersion data\cite{statler99};
the points show the PN velocity data.
The background shading shows the projection of the best-fit model,
multiplied by the stellar luminosity at each radius;
the dotted line shows its dispersion.

 \includegraphics[width=\textwidth]{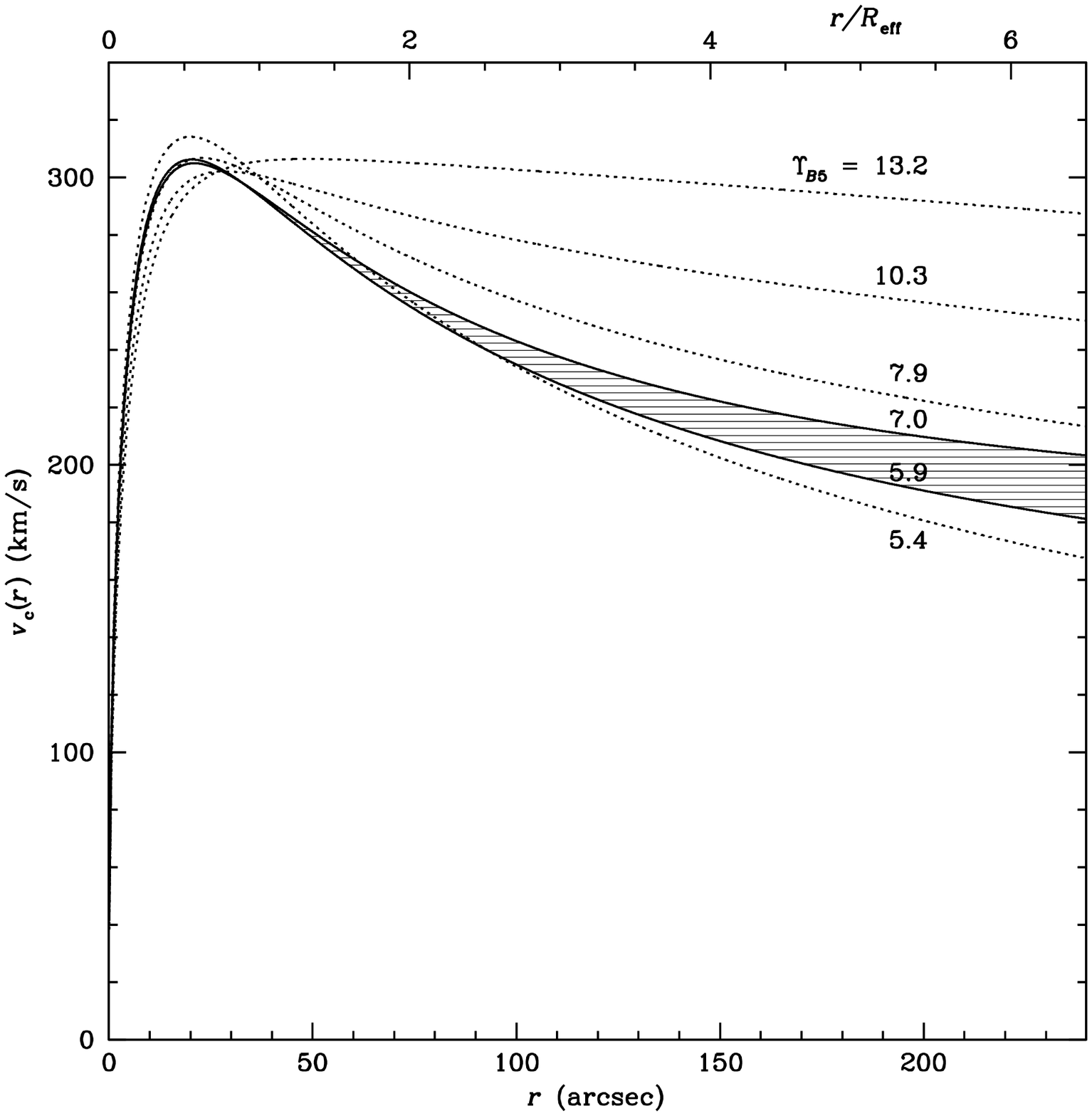}
\noindent {\bf Fig. A2.}
Intrinsic circular velocity profile with radius, $v_{\rm c}(r) \equiv \sqrt{G M(r)/r}$, of NGC~3379.
The solid lines and shaded area show the region permitted by orbit modeling.
The dotted lines show models which are ruled out at the 1-$\sigma$ level:
the bottom one shows a constant mass-to-light ratio ($M/L$) model,
the upper three show models with more dominant dark halos.
Each model is labeled with
its {\it B}-band mass-to-light ratio (in solar units) at 5~effective radii.
Thus we see that the shape of $v_{\rm c}(r)$ for a 
constant $M/L$ galaxy falls slightly outside the allowed range.

\end{document}